\documentclass[english,runningheads,a4paper]{llncs}[2018/03/10]
\usepackage[ngerman,main=english]{babel}
\addto\extrasenglish{\languageshorthands{ngerman}\useshorthands{"}}
\usepackage{regexpatch}
\makeatletter
\edef\switcht@albion{%
  \relax\unexpanded\expandafter{\switcht@albion}%
}
\xpatchcmd*{\switcht@albion}{ \def}{\def}{}{}
\xpatchcmd{\switcht@albion}{\relax}{}{}{}
\edef\switcht@deutsch{%
  \relax\unexpanded\expandafter{\switcht@deutsch}%
}
\xpatchcmd*{\switcht@deutsch}{ \def}{\def}{}{}
\xpatchcmd{\switcht@deutsch}{\relax}{}{}{}
\edef\switcht@francais{%
  \relax\unexpanded\expandafter{\switcht@francais}%
}
\xpatchcmd*{\switcht@francais}{ \def}{\def}{}{}
\xpatchcmd{\switcht@francais}{\relax}{}{}{}
\makeatother

\usepackage{ifluatex}
\ifluatex
  \usepackage{fontspec}
  \usepackage[english]{selnolig}
\fi

\iftrue 
  \ifluatex
  \else
    \usepackage[%
      rm={oldstyle=false,proportional=true},%
      sf={oldstyle=false,proportional=true},%
      tt={oldstyle=false,proportional=true,variable=false},%
      qt=false%
    ]{cfr-lm}
  \fi
\else
  \ifluatex
  \else
    \usepackage{newtxtext}
    \usepackage{newtxmath}
    \usepackage[zerostyle=b,scaled=.9]{newtxtt}
  \fi
\fi

\ifluatex
\else
  \usepackage[T1]{fontenc}
  \usepackage[utf8]{inputenc} 
\fi

\usepackage{graphicx}

\usepackage{upquote}
\usepackage{booktabs}
\usepackage{paralist}
\usepackage{csquotes}

\usepackage{textcmds}

\RequirePackage[%
  babel,%
  final,%
  expansion=alltext,%
  protrusion=alltext-nott]{microtype}%
\DisableLigatures{encoding = T1, family = tt* }

\usepackage{url}
\makeatletter
\g@addto@macro{\UrlBreaks}{\UrlOrds}
\makeatother

\usepackage{xcolor}

\usepackage{listings}
\renewcommand{\lstlistingname}{List.}
\usepackage{chngcntr}
\AtBeginDocument{\counterwithout{lstlisting}{section}}

\usepackage{pdfcomment}
\usepackage[%
  square,        
  comma,         
  numbers,       
  sort&compress, 
]{natbib}

\ifluatex
\else
  \SetExpansion
  [ context = sloppy,
    stretch = 30,
    shrink = 60,
    step = 5 ]
  { encoding = {OT1,T1,TS1} }
  { }
\fi

\usepackage{stfloats}
\fnbelowfloat

\usepackage{hyperref}
\hypersetup{hidelinks,
  colorlinks=true,
  allcolors=black,
  pdfstartview=Fit,
  breaklinks=true}
\usepackage[all]{hypcap}

\usepackage[group-four-digits,per-mode=fraction]{siunitx}

\usepackage[capitalise,nameinlink]{cleveref}
\usepackage{iflang}
\IfLanguageName{ngerman}{
  \crefname{table}{Tab.}{Tab.}
  \Crefname{table}{Tabelle}{Tabellen}
  \crefname{figure}{\figurename}{\figurename}
  \Crefname{figure}{Abbildungen}{Abbildungen}
  \crefname{equation}{Gleichung}{Gleichungen}
  \Crefname{equation}{Gleichung}{Gleichungen}
  \crefname{listing}{\lstlistingname}{\lstlistingname}
  \Crefname{listing}{Listing}{Listings}
  \crefname{section}{Abschnitt}{Abschnitte}
  \Crefname{section}{Abschnitt}{Abschnitte}
  \crefname{paragraph}{Abschnitt}{Abschnitte}
  \Crefname{paragraph}{Abschnitt}{Abschnitte}
  \crefname{subparagraph}{Abschnitt}{Abschnitte}
  \Crefname{subparagraph}{Abschnitt}{Abschnitte}
}{
  \crefname{section}{Sect.}{Sect.}
  \Crefname{section}{Section}{Sections}
  \crefname{listing}{\lstlistingname}{\lstlistingname}
  \Crefname{listing}{Listing}{Listings}
}

\usepackage{xspace}

\newcommand{\ie}{i.\,e.,\ }

\DeclareFontFamily{U}{MnSymbolC}{}
\DeclareSymbolFont{MnSyC}{U}{MnSymbolC}{m}{n}
\DeclareFontShape{U}{MnSymbolC}{m}{n}{
  <-6>    MnSymbolC5
  <6-7>   MnSymbolC6
  <7-8>   MnSymbolC7
  <8-9>   MnSymbolC8
  <9-10>  MnSymbolC9
  <10-12> MnSymbolC10
  <12->   MnSymbolC12%
}{}
\DeclareMathSymbol{\powerset}{\mathord}{MnSyC}{180}

\ifluatex
\else
  \input glyphtounicode
\fi

\usepackage[math]{blindtext}
\usepackage{mwe}

\begin{document}
\sloppy

\title{Towards Compliant Data Management Systems for Healthcare ML}
\titlerunning{Towards Compliant Data Management Systems}


\author{
    Goutham Ramakrishnan\thanks{The work was performed during an internship at Microsoft Research, Cambridge, UK}\inst{1} \and 
    Aditya Nori\inst{2} \and
    Hannah Murfet\inst{2} \and 
    Pashmina Cameron\inst{2}
}
\authorrunning{Ramakrishnan et al.}
\institute{
    University of Wisconsin--Madison \and
    Microsoft Research Cambridge
}

\maketitle

\begin{abstract}
The increasing popularity of machine learning approaches and the rising awareness of data protection and data privacy presents an opportunity to build truly secure and trustworthy healthcare systems. Regulations such as GDPR and HIPAA present broad guidelines and frameworks, but the implementation can present technical challenges. 
Compliant data management systems require enforcement of a number of technical and administrative safeguards. While policies can be set for both safeguards there is limited availability to understand compliance in real time.
Increasingly, machine learning practitioners are becoming aware of the importance of keeping track of sensitive data. 
With sensitivity over personally identifiable, health or commercially sensitive information there would be value in understanding assessment of the flow of data in a more dynamic fashion. 
We review how data flows within machine learning projects in healthcare from source to storage to use in training algorithms and beyond. Based on this, we design engineering specifications and solutions for versioning of data. Our objective is to design tools to detect and track sensitive data across machines and users across the life cycle of a project, prioritizing efficiency, consistency and ease of use. We build a prototype of the solution that demonstrates the difficulties in this domain. Together, these represent first efforts towards building a compliant data management system for healthcare machine learning projects.
\end{abstract}


\section{Introduction}\label{sec:intro}
\begin{figure}[t!]
  \center
  \includegraphics[width=1\linewidth]{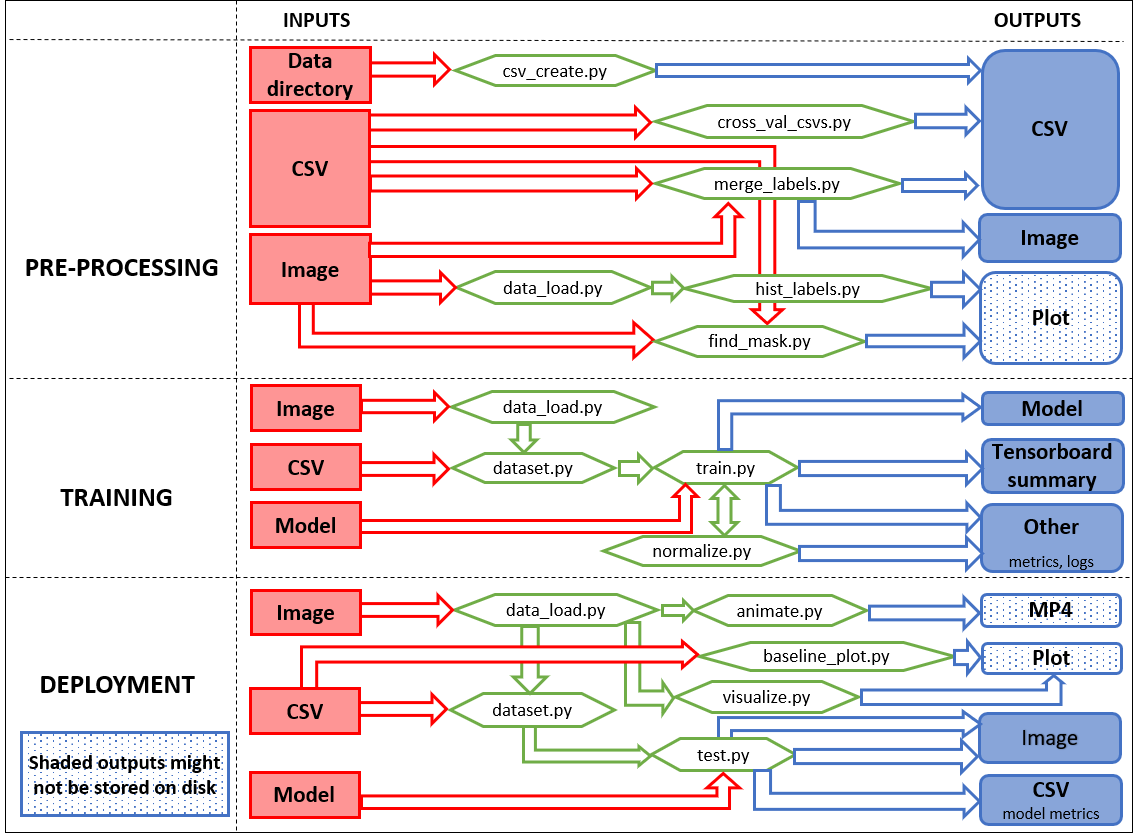}
  \caption{The data-flow across the stages of a machine learning pipeline for healthcare}
  \label{fig:afnet}
\end{figure}

Data-driven machine learning approaches have grown in prominence over the last decade, including in cross-disciplinary research areas such as healthcare. 
In the healthcare domain, the data is often of a sensitive and confidential nature, thus requiring careful data protection and privacy practices by the data users.
For example, images of MRI scans obtained from hospitals, which are used to train machine learning models for cancer diagnosis (see Figure \ref{fig:afnet}), often contain personally identifiable information (PII) or Protected Health Information (PHI) at source. 
The use of such sensitive data necessitates extreme care to ensure compliance with local data protection laws and
any contractual obligations. Failure to do so may have damaging consequences for the patients, for example one scenario could be a situation where a group of individual's health records leaked, identity information stolen, and consequently they become a target of phishing attacks. Organizations who act as data controllers may face consequences such as loss of trust and reputation, loss of business, hefty fines\cite{healthcare-gdpr} or even lawsuits \cite{dpact}. Contractual obligations often take the form of a data sharing agreement between the data sharer and the research institution. Often these  data sharing agreements require constraints \cite{limited-data-sets} including how the data will be used, for what purposes, for use by specific individuals, require deletion after a given retention period, to not attempt to re-identify individuals and prohibit further disclosure.

With the increasing prevalence of data-driven approaches, regulatory bodies across the world have already or are in the process of enacting strict data protection and privacy laws. 
General Data Protection Regulation (GDPR)\cite{GDPR} is a prominent example of such a law, which provides safeguards for citizens of the European Union. HIPAA \cite{hipaa} is an example of data privacy and security regulation from the United States.  
GDPR enforces strict regulations on the collectors and processors of data and imposes heavy penalties upon their violation. 
Organizations usually appoint a Data Protection Officer (DPO), who can help monitor internal data access and advise on matters relating to data protection, for ensuring accountability. 
From our review, there are limited software tools which are specifically designed to help and improve the work efficiency of DPOs and privacy officers that support the full data lifecycle.

Tracking the flow (and copies) of data in real time, before a project starts, during the project and after the project has concluded can be challenging, and currently there are few tools to support data tracking over the entire project life cycle. Version control tools for source code are well understood, but the tools required for data versioning are still in the nascent stage. Compliance in healthcare requires a good data versioning system as a minimum, but has needs that beyond this. Data flow within healthcare needs to be tracked not only across different versions, but also across different permission levels and different data classifications (e.g. anonymous vs. pseudonymous).
The challenges associated with developing such tools stem from the inherent complexity of modern project workflows. 
\begin{itemize}
  \item Each project may have multiple researchers working on it, each researcher may have multiple machines on which they work. 
  \item Each machine may also be used by more than one researcher. 
  \item Each project may have multiple data sets and each data set may contain multiple sensitive data elements.
  \item A project or a study may conclude, which may lead to a need to delete all copies of the data used for such a study
  \item The permission to use certain data sets may be time-limited, leading to a need to track and delete all copies of the data set including derivatives
\end{itemize} 
Keeping these in mind, our objective is to design tools to detect and track sensitive data across machines and users across the life cycle of a project, prioritizing efficiency, consistency and ease of use. Specifically, at various stages of a project, the permissions to access, modify and delete data may vary, so we need tools to periodically (at some pre-specified frequency) assess the compliance of data storage and usage to the relevant regulations. 

We aim to build tools to \textit{enforce} compliance while keeping the overhead minimal. More often than not, there is a deep desire to comply to data protection rules, but these rules can be quite complex and may change over time. Integrating the data protection rules within the machine learning workflows for healthcare makes using automated tools for tracking data movements makes it easier for researchers to build trustworthy systems. 
Healthcare ML practitioners are the primary data feeders to the tool and the DPO or privacy officer is the primary consumer of the tool in order to monitor data compliance. As adoption of the tool grows, more stringent compliance requirements can be monitored and adhered to within the tool. 

\textit{Complete and provably correct enforcement} of data compliance is a very difficult task, and as yet an unsolved problem. There have been some attempts to solve a part of the problem \cite{wang2019data} using formal policy language, but such tools largely use static policies (\ie they only inspect the source code) to perform verification; compliance with data protection rules requires software validation as well. Each project has different requirements, each machine learning system has different vulnerabilities and each region and type of data has different levels of sensitivity associated with it. 
Discovering and mitigating vulnerabilities and data leakage (or model leakage) within machine learning \cite{dl_vul,badnets,nn_security} (or indeed even a given software system and hardware platform \cite{foreshadow,spectre}) is a very active area of research. 
Even if all three of these aspects were fully understood, any system that imposes full `surveillance' upon those working with the data would indeed feel over burdensome to a researcher new to working in this area. 
Recently, there has been an increased focus on privacy regulations, and it will take some time for users to fully appreciate and accustom themselves with the changing landscape. 
Thus, in order to encourage uptake of a new tool, changes made to existing systems have to be gradual and incremental, always keeping in mind the human factor.  
Without formal verification guarantees, we have no way of automatically ensuring the compliance of the outputs generated by the scripts used to process the data. 

We envision a data management system that is built specifically with the aim of enforcing and demonstrating compliance to data protection regulations. 
The design and implementation of such a system must facilitate ease of use and prove useful across different roles in an organization. 
While it must help individual researchers and developers keep track of their data, it must also function in a broader scope to enable managers and DPOs/ privacy officers to track compliance across projects and teams at the organizational level.

The data management tool must comply with the following high-level specifications, in order to ensure data protection (privacy and security):
\begin{description}
  \item[\textit{Data discovery:}] Find all data with potentially sensitive data across all storage mechanisms, all relevant file formats, and machines for all users. Determine the level of sensitivity of the data. Scan the storage systems at a user-specified frequency to facilitate automated monitoring of data. 
  \item[\textit{Data versioning:}] Detect all file changes since last scan, including changes to contents, metadata or location of the file. Detect if any files have been deleted or created. Use hashing to compare contents of a file. Log this information using a globally unique identifier such that each file, the machine on which the files resides and the user of the data can be identified uniquely. 
  \item[\textit{Storage and review:}] Store the information gathered from the scans in a database. The database schema must be such that an intuitive interface for reviewing the stored logs can be implemented efficiently. Ensure recentness and consistency of logs. Allow a concept of staleness in the database to indicate out-of-date information in case the scans fail for some reason. 
\end{description}

As a first step towards building such a tool, we have built a prototype system with the following fine-grained software specification: 
\vspace{-\topsep}
\begin{enumerate}[(1)]
\item Each sensitive file should be tracked by the system. Each file is uniquely identified by it's filepath and SHA-256 hash. 
\begin{itemize}
  \item When a file is moved or copied, a new file is considered to have been created due to changes to it's filepath. 
  \item When the contents of a file are changed, a file is considered to be modified due to changes to it's SHA-256 hash. 
\end{itemize}
\item Researchers handling the data specify the root folders of all locations (including all machines) where sensitive data is stored. The completeness of the scans done by the compliance tool relies on researchers taking care to provide a complete list of root folders. Subfolders and compressed folders are checked automatically. 
\item File formats such as DICOM contain image pixel data and patient information as metadata in a single bundle. It may be important to track modifications in the pixel data and metadata independently (using hash values). 
\item Metadata such as `Last Scanned' and `Last Modified' are useful to check if the logged information is up-to-date. 
\item If sensitive data is read and processed by a program, and some derivative data is created, the derivative data is assumed to be sensitive in the absence of any further information provided by the researchers. In a future version, researchers may be able to whitelist certain programs that they know do not produce any sensitive derived data. Additionally, \textit{formal verification} tools such as \cite{ironclad, formal_dnn} may be used to automatically determine whether certain programs conform to their original software specification (and thus provide guarantees on the sensitivity of the derived data), but this is beyond the scope of this work. 
\end{enumerate}



\section{Why data versioning is not enough}
\label{sec:relatedwork}
Version control for source code is a well-studied problem in the domain of software engineering and several implementations are available for use, with git\cite{git} being one of the most preferred tools used by developers. 
Machine learning research requires version control for data in addition to source code. 
With the increasing emphasis on reproducible research in machine learning over the last few years, several tools which provide the facility of data versioning have emerged. 

Git-LFS\cite{git-lfs} provides storage functionality for large data files along with seamless integration with git. It replaces large files with text pointers inside git, while the data itself is stored on a remote server. However a major shortcoming is that the files are stored as blobs and thus the data files cannot be diffed. 
Pachyderm\cite{pachyderm} offers a comprehensive data versioning system, with repositories for data, flexible data pipelines and support for cloud-based computing. However it is a heavyweight software, and the data needs to be copied in and out of the Pachyderm cluster to use it. 
Quilt\cite{quilt} offers a simple model that represents data as python packages. However, it may not scale well to large data inputs and its usage-APIs are implemented only in python and R. 
DVC\cite{dvc} provides the functionality of data dependency tracking along with git integration. It is an open-source tool, easy to learn and run. 
MissingLink\cite{missinglink} offers an end-to-end management of machine learning projects, with version control for data being a key feature. This requires users to upload data on to missinglink.ai servers. Some commercial solutions that are not tailored to a machine learning workflow do exist \cite{filerskeepers}, but these ordinarily don't cater to machine learning workflows where high-throughput data access for training, fine-tuning and debugging models is required. In some cases, the contractual restrictions may also require the data to be hosted within certain institutional boundaries, which may mean that such a tool cannot be used if the data is not stored on premise. 

However, the use of the above tools on sensitive data in the healthcare domain presents several challenges from the compliance perspective. Sensitive data cannot be uploaded to third-party servers, first because the data may leak, or be tampered with, and second because the servers may well be in a different geographic region from the data, both of which may not be acceptable from a compliance point of view. File formats such as DICOM and NIFTII constitute a bundle of patient information along with pixel data, it may be important to individually keep track of changes in both. Moreover, the sensitive data files which may be created over the course of the project cycle need to be automatically detected and tracked. The unique constraints imposed by the sensitivity of the data and the inadequacy of the existing tools for adherence to compliance (as opposed to mere enforcement) motivates the design and implementation of a new data management system for compliance.

\section{Methodology}
\label{sec:methodology}

Sensitive data files may be used/created across the various stages of a typical machine learning pipeline. 
In the data pre-processing stage, the raw data is cleaned, transformed and converted for consumption by the model. 
Often, it becomes necessary for researchers to download the raw/transformed training data for debugging purposes. 
The training, debugging and testing phases of the project may also lead to creation of sensitive data files. 
The complexity of data flows of an actual machine learning project are shown in the flowchart in Figure \ref{fig:afnet}.  
A compliant data management system should give users the ability to track and monitor sensitive data across machines across the project cycle, thus making it easy for them to be compliant. 
We visualize such a system to have three key components: data discovery, data versioning and data review.

\subsection{Data Discovery} 
A key prerequisite for any system that tracks data compliance is the ability to find all files containing sensitive data. Asking the user to specify which files are sensitive rather puts the onus on to the ML researcher. The specific files that contain sensitive data may also change over the course of the project. Often sensitive data can be isolated to one or more specialized file formats for a particular project (for example, a project using MRI scan images may restrict itself to just DICOM files). 
However, a simple directory-walk or search based on file extensions is not adequate for finding all sensitive files. 
For completeness of the file discovery, in addition to searching for specified file extensions, it is necessary to account for the following: 
\vspace{-\topsep}
\begin{enumerate}[(a)]
\item Files may be present in a compressed format (e.g. as a `gz' or inside a `zip')
\item Files may be present inside temporary folders (and therefore not have the same file extension)
\item Files may be present under a different extension as a result of bugs in code generating them or deliberate attempt to fool file discovery. In such cases, it becomes necessary to check the file signature (usually by reading the initial few bytes of the file) to determine the file format. It may also be useful to gather additional information about the files. For example, DICOM files may not have the `.dcm' extension, may or may not have a well-formed preamble and may be stored in little or big endian.  
\item There may also be files in a system that present themselves as one of our specified formats, but are in fact just text or image files. The system must be able to deal with these files accordingly. For example, a `.dcm' file may be renamed as `.jpg', a `.jpg' may be renamed as a `.dcm'. When configured to look for `.dcm' file, all files that are truly `.dcm', whatever the extension must be identified and tracked by our compliance checker.    
\end{enumerate}

\noindent Depending on the properties of any particular file format, additional steps may be necessary to ensure the exhaustiveness of the file discovery.

\subsection{Data Versioning}
Across the life-cycle of a machine learning project, new data files may be created and old data files may be modified, moved or deleted. 
From a compliance perspective, it is imperative that we are able to track such events. 
We propose a file versioning system built according to the following guidelines: 
\vspace{-\topsep}
\begin{enumerate}[(a)]
\item Each sensitive file is tracked by considering its filepath to be invariant. When it is moved or copied, a new file is considered to have been created.  
\item Changes in the file (and therefore new versions) are detected by comparing with the SHA-256 hash values of the file contents. 
\item File formats such as DICOM contain image pixel data and patient information as metadata in a single bundle. It may be important to track modifications in the pixel data and metadata independently (using hash values). 
\item Metadata such as `Last Scanned' and `Last Modified' are useful to check if the logged information is up-to-date. 
\end{enumerate}
The information gathered from the `Compliance Checker Scan' (i.e. data discovery followed by data versioning) must be stored in the back end. 

\subsection{Data Review Interface}
The information logged by the data discovery and versioning stages must be easily available for review, in an intuitive interface. 
Ideally: 
\vspace{-\topsep}
\begin{enumerate}[(a)]
\item The interface must automatically retrieve the latest information from the backend. 
\item It must allow the DPO or privacy officer to filter and view information pertaining to specific fields (such as machine, user, scan time, file format, etc.)
\item It must allow the DPO or privacy officer to check for the recentness of the information in the logs. 
\item A utility that allows the DPO or privacy officer to email the user of a given machine in order to remind them to run the scan would be very useful as well. 
\end{enumerate}
Depending on the nature of the sensitive data and the requirements of the DPO or privacy officer, the review interface may require further functionalities.

\section{Implementation}
\label{sec:implementation}
We describe our implementation of a software prototype for the tool, the `Compliance Checker', which uses Python3.7, Azure SQL and PowerBI.

\subsection{CompliCheck: A python package}
The Compliance Checker software is developed as a python package, the key features of which are described below: 
\vspace{-\topsep}
\begin{itemize}
\item The code structured using the principles of object oriented programming, and is thus easy to understand and build upon. 
\item The file discovery component of the package has custom file discoverers for DICOM and NIFTII file formats. The modular structure of the code allows one to add more custom file discoverers easily. It also searches inside compressed files and folders. 
\item The file versioning component of the package primarily acts as an interface of communication with the Azure SQL database, where the results of the compliance scans are logged. 
\item The package can be easily installed using the `setup.py' script. 
\end{itemize}

\begin{figure}[t!]
  \center
  \includegraphics[width=0.7\linewidth]{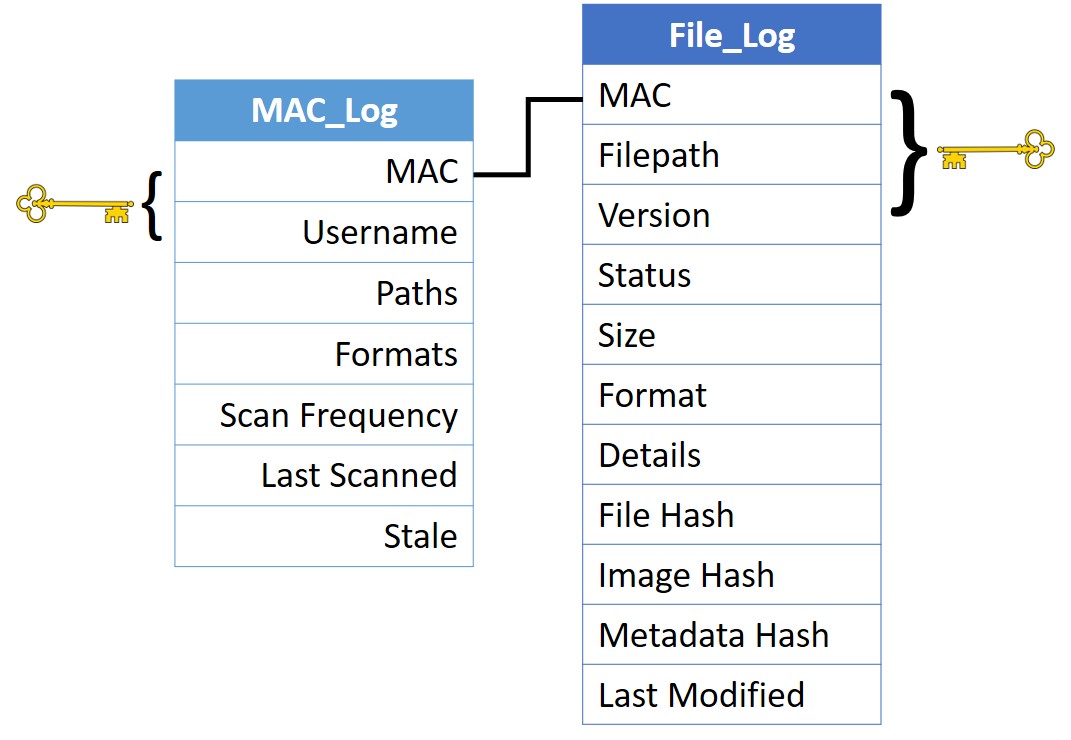}
  \caption{Database Schema used by the Compliance Checker}
  \label{fig:schema}
\end{figure}

\subsection{Database Design}
The schema of the Azure SQL database used is shown in Figure \ref{fig:schema}. 
The database is structured as two tables, File\_Log and MAC\_log, which are described below:
\begin{enumerate}[(a)]
\item \textbf{MAC\_Log:}
This table contains one row per user (identified by `Username') per machine (identified by MAC address). Each entry specifies the sensitive file formats to scan for, and the filepaths in which to run the scan. It is important to verify the recentness/freshness of the scans, thus we maintain the `Stale' field. The `Scan Frequency' specifies the ideal frequency of the compliance scan, and the `Last Scanned' stores the datetime corresponding to the last time the scan was run. An entry is considered `Stale' if a time greater than the `Scan Frequency' has elapsed since `Last Scanned'. Note that storing this information allows us to update the values of the `Stale' column in a self-sufficient manner, without needing information from the File\_Log.  
 \item \textbf{File\_Log:}
This table contains the file versioning information for every file scanned on every machine. Note that we do not store the `Username' field in this table, to avoid duplicate records in case of shared folders on machines. The `Version' is an integer, and the `Status' specifies whether an entry is `Latest Version', `Old Version' or `Deleted'. The `Last Modified' field stores the scan datetime of the scan when the version (i.e. change in file) was detected. The other fields in the table are self-evident, and have been justified in Section \ref{sec:methodology}.    
\end{enumerate}

\subsection{Running the Compliance Checker}
When running the compliance checker for the first time for a new machine-user combination, an entry must be created in the MAC\_Log table first. 
When the `Run' operation is called, the paths and formats to be scanned are retrieved from the MAC\_Log.
The compliance scan begins and the results are stored in the File\_Log table. 
Note that the compliance scan is an atomic operation, i.e. the scan either succeeds or fails in its entirety. 
The changes to the backend database are only committed at the end of the scan, thus ensuring consistency.
A `Paths', `Formats' and `Scan Frequency' of a pre-existing entry in the MAC\_Log can be modified if necessary. 
To ensure recentness/freshness of the `Stale' column in the MAC\_Log, all entries are updated at the end of a compliance scan on any machine.  
All of the above operations can be achieved through running the program with appropriate arguments. 

Note that our implementation of data versioning is `storage-free'. Therefore, it only detects new versions of files and does not offer the ability to revert to old versions. Development of this feature is left for future work.  

\begin{figure}[t!]
  \includegraphics[width=\linewidth]{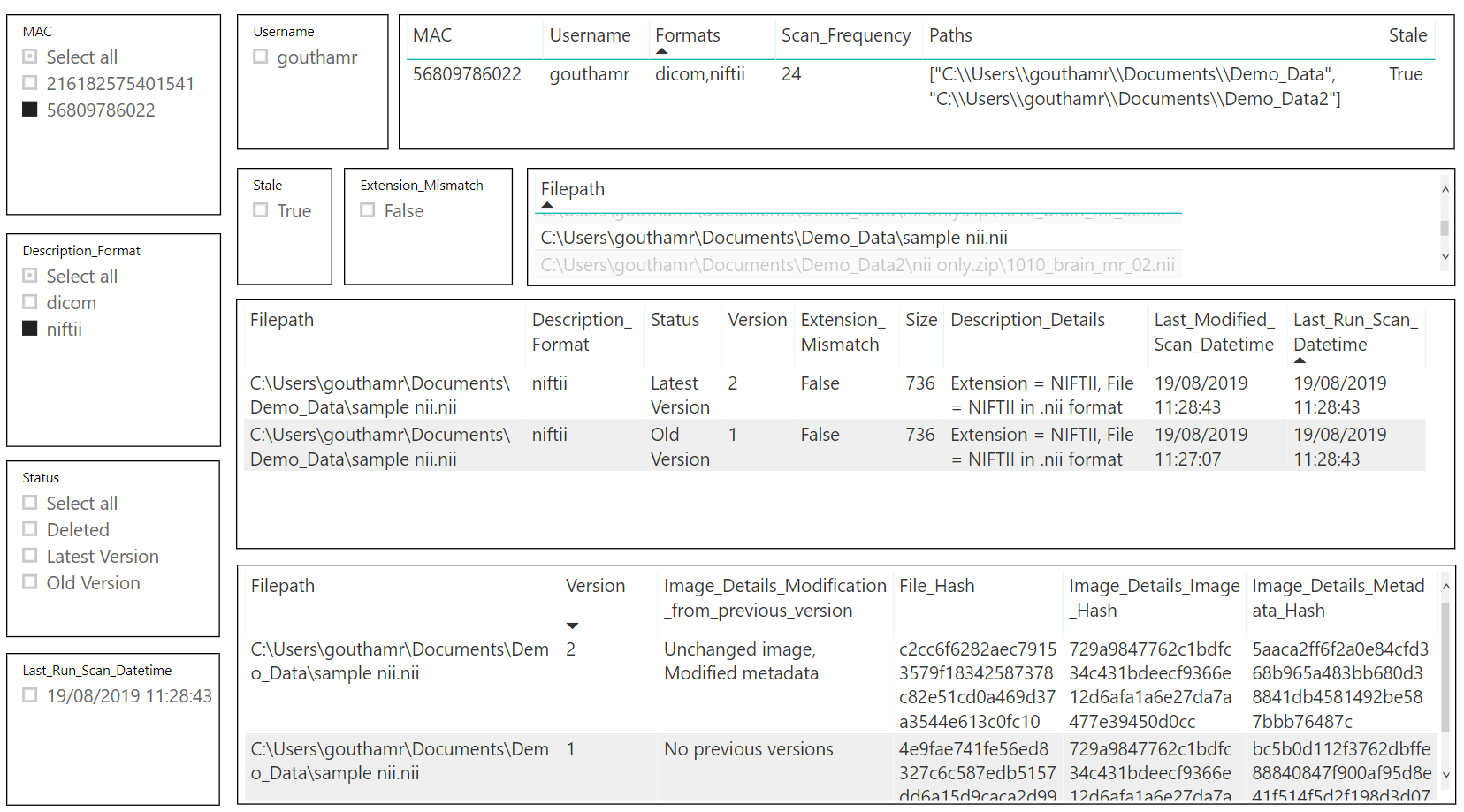}
  \caption{PowerBI Report for Compliance Review. The deviations from previous versions of the data can be reviewed by the DPO or privacy officer using this interface.}
  \label{fig:powerbi}
\end{figure}

\subsection{Deployment and Automation}
The development of the software as a python package enables us to deploy the tool across machines easily. 
In addition, to ensure regularity of scanning, it is important to automate the running of the tool on machines. 
On Linux machines, this can be implemented as a cron job. 
On Windows machines, automation can be achieved by scheduling a task to run periodically. 
Automatic scans are especially useful and important when working with virtual machines/cloud computing. 
To this end, we have developed batch/bash scripts for use by system administrators to install the software and setup the automation on machines.   

\subsection{PowerBI Interface}
As a review interface, we have created a PowerBI report for viewing the results of the compliance checker program (as shown in Figure \ref{fig:powerbi}). 
The interface allows the DPO or privacy officer to view the results of the compliance scans across machines, filtered by format, scan-time, version and staleness. 
It is directly linked to the Azure SQL database to retrieve the latest data. This is a viewing tool only, at the moment. In future, a more comprehensive user interface could provide additional functionality such as adding file formats to be checked or sending reminders to users to run the compliance checker tool on a particular machine if stale data is detected.

\section{Conclusion}
Compliance with data privacy and protection laws are a priority for modern organizations, thus motivating the need for a data management system built specifically for the purpose of facilitating adherence to compliance regulations. Even when entities have the best intentions, proving compliant behaviour is extremely hard. Therefore, at some level, collaborations in healthcare (or other spaces that includes handling sensitive or personal data) are based on trust. However, with recent high-profile cases have proven that implicit promise of trust is no longer sufficient and we need to put system level checks in place to ensure and verify good behaviour. 
In this paper, we have reported on ongoing work towards the design of tools to that end and present a prototype implementation of the software. 

Future work will explore (a) a more nuanced methodology for discovering sensitive data, for example, discovering PII data, anonymised vs non-anonymised data, etc.; (b) building a fully-fledged data version control system with compliance in mind rather than trying to add compliance as an after-thought; (c) building a dedicated software front-end for review that also allows the DPO or privacy officer to send reminders to project team members. In a future version, one could also envisage a customizable definition of data versioning for each application. The changes that constitute a new version may not be the same for each application and it would be desirable to track the attributes that matter most for each use case.

\renewcommand{\bibsection}{\section*{References}} 
\bibliographystyle{splncsnat}
\begingroup
  \ifluatex
  \else
    \microtypecontext{expansion=sloppy}
  \fi
  \small 
  \bibliography{paper.bib}

\begin{thebibliography}{20}
\providecommand{\natexlab}[1]{#1}
\providecommand{\url}[1]{\texttt{#1}}
\providecommand{\urlprefix}{}

\bibitem[{Acts(2018)}]{dpact}
Acts, U.P.G.: {D}ata {P}rotection {A}ct.
\newblock \url{https://www.legislation.gov.uk/ukpga/2018/12/contents/enacted}
  (2018)

\bibitem[{Agustin(2019)}]{dl_vul}
Agustin, C.E.S.: Mitigating deep learning vulnerabilities from adversarial
  examples attack in the cybersecurity domain.
\newblock CoRR abs/1905.03517 (2019),
  \urlprefix\url{http://arxiv.org/abs/1905.03517}

\bibitem[{DVC(2017)}]{dvc}
DVC: {Machine Learning Version Control System}.
\newblock \url{https://dvc.org} (2017)

\bibitem[{{European Commission}(2016)}]{GDPR}
{European Commission}: Regulation ({EU}) 2016/679 of the european parliament
  and of the council({GDPR}).
\newblock \url{https://eur-lex.europa.eu/eli/reg/2016/679/oj} (2016)

\bibitem[{FilersKeepers(????)}]{filerskeepers}
FilersKeepers: {filerskeepers}.
\newblock \url{https://www.filerskeepers.co/} (????)

\bibitem[{Foreshadow(2018)}]{foreshadow}
Foreshadow: {Understanding L1 Terminal Fault aka Foreshadow: What you need to
  know}.
\newblock
  \url{https://www.redhat.com/en/blog/understanding-l1-terminal-fault-aka-foreshadow-what-you-need-know}
  (2018)

\bibitem[{{Fred Hutch extranet}(????)}]{limited-data-sets}
{Fred Hutch extranet}: {PHI - Difference Between De-Identified and Limited Data
  Sets}.
\newblock \url{https://extranet.fredhutch.org/en/u/hdc/data.html} (????)

\bibitem[{Git(2005)}]{git}
Git: {Git}.
\newblock \url{https://git-scm.com} (2005)

\bibitem[{Git LFS(2015)}]{git-lfs}
Git LFS: {Git Large File Storage}.
\newblock \url{https://git-lfs.github.com} (2015)

\bibitem[{Gu et~al.(2017)Gu, Dolan{-}Gavitt, and Garg}]{badnets}
Gu, T., Dolan{-}Gavitt, B., Garg, S.: Badnets: Identifying vulnerabilities in
  the machine learning model supply chain.
\newblock CoRR abs/1708.06733 (2017),
  \urlprefix\url{http://arxiv.org/abs/1708.06733}

\bibitem[{Hawblitzel et~al.(2014)Hawblitzel, Howell, Lorch, Narayan, Parno,
  Zhang, and Zill}]{ironclad}
Hawblitzel, C., Howell, J., Lorch, J.R., Narayan, A., Parno, B., Zhang, D.,
  Zill, B.: Ironclad apps: End-to-end security via automated full-system
  verification.
\newblock In: 11th {USENIX} Symposium on Operating Systems Design and
  Implementation ({OSDI} 14). pp. 165--181. {USENIX} Association, Broomfield,
  CO (Oct 2014),
  \urlprefix\url{https://www.usenix.org/conference/osdi14/technical-sessions/presentation/hawblitzel}

\bibitem[{MissingLink(2016)}]{missinglink}
MissingLink: {Missinglink.ai}.
\newblock \url{https://missinglink.ai} (2016)

\bibitem[{Neural Networks(2019)}]{nn_security}
Neural Networks: {Security Vulnerabilities of Neural Networks}.
\newblock
  \url{https://towardsdatascience.com/hacking-neural-networks-2b9f461ffe0b}
  (2019)

\bibitem[{Pachyderm(2014)}]{pachyderm}
Pachyderm: {Pachyderm - Scalable, Reproducible Data Science}.
\newblock \url{https://www.pachyderm.io} (2014)

\bibitem[{Quilt(2017)}]{quilt}
Quilt: {Quilt Data Inc}.
\newblock \url{https://quiltdata.com} (2017)

\bibitem[{Register(2018)}]{healthcare-gdpr}
Register, G.: Healthcare sector: How to comply with gdpr?
\newblock \url{https://www.gdprregister.eu/gdpr/healthcare-sector-gdpr/} (2018)

\bibitem[{Seshia et~al.(2018)Seshia, Desai, Dreossi, Fremont, Ghosh, Kim,
  Shivakumar, Vazquez-Chanlatte, and Yue}]{formal_dnn}
Seshia, S.A., Desai, A., Dreossi, T., Fremont, D., Ghosh, S., Kim, E.,
  Shivakumar, S., Vazquez-Chanlatte, M., Yue, X.: Formal specification for deep
  neural networks.
\newblock Tech. Rep. UCB/EECS-2018-25, EECS Department, University of
  California, Berkeley (May 2018),
  \urlprefix\url{http://www2.eecs.berkeley.edu/Pubs/TechRpts/2018/EECS-2018-25.html}

\bibitem[{Spectre(2019)}]{spectre}
Spectre: {How the Spectre and Meltdown Hacks Really Worked}.
\newblock
  \url{https://spectrum.ieee.org/computing/hardware/how-the-spectre-and-meltdown-hacks-really-worked}
  (2019)

\bibitem[{{U.S. Department of Health and Human Services}(1996)}]{hipaa}
{U.S. Department of Health and Human Services}: {The Health Insurance
  Portability and Accountability Act (HIPAA)}.
\newblock
  \url{https://www.hhs.gov/hipaa/for-professionals/security/laws-regulations/index.html}
  (1996)

\bibitem[{Wang et~al.(2019)Wang, Near, Somani, Gao, Low, Dao, and
  Song}]{wang2019data}
Wang, L., Near, J.P., Somani, N., Gao, P., Low, A., Dao, D., Song, D.: Data
  capsule: A new paradigm for automatic compliance with data privacy
  regulations (2019)

\end{thebibliography}
\endgroup

\ \\
\end{document}